\documentclass[12pt,a4paper,twoside,english]{article}
\usepackage[T1]{fontenc}
\usepackage[latin1]{inputenc}
\usepackage{amsbsy}
\usepackage{graphicx}



\usepackage{babel}
\begin{document}
\title{Efficient cooling by ferroelectric or ferromagnetic hysteresis loops}
\author{{\normalsize{}M. Apostol }\\
{\normalsize{}Department of Theoretical Physics, Institute of Atomic
Physics, }\\
{\normalsize{}Magurele-Bucharest MG-6, POBox MG-35, Romania }\\
{\normalsize{}email: apoma@theory.nipne.ro}}
\date{{}}

\maketitle
Orchid: 0000-0002-9990-9390

\relax
\begin{abstract}
An efficient cooling effect is put forward, by means of external electric
or magnetic fields along hysteresis loops. A simplified model of hysteresis
is used for numerical illustration. The model is based upon a second-order
expansion of the energy in powers of polarization and external field.
The electrocaloric effect along hysteresis loops is discussed for
comparison.
\end{abstract}
\relax


\section{Introduction. Hysteresis}

As it is well known, hysteresis is an old subject, both for ferromagnets
and ferroelectrics (see, for example, Refs. \cite{key-1}-\cite{key-4}).
On the other hand, the use of such materials for cooling is a promising
possibility, which enjoys much interest.\cite{key-5}-\cite{key-10}
We suggest in this paper an efficient cooling effect, by means of
external electric or magnetic fields along hysteresis loops. For numerical
illustration we use a simplified model of hysteresis. The results
are valid both for ferroelectrics and ferromagnets; we specialize
the presentation to ferroelectrics.

\begin{figure}
\begin{centering}
\includegraphics[clip]{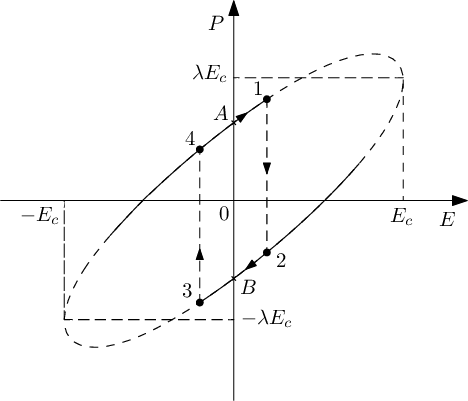}
\par\end{centering}
\caption{Hysteresis cycle.\label{fig:Fig.1}}
\end{figure}

Let us consider an isolated ferroelectric below its critical temperature,
which exhibits domains and hysteresis. We adopt a simplified model,
characterized by a local polarization $\boldsymbol{P}_{0}$ and an
energy density $\mathcal{E}(P_{0})$ associated to this polarization,
which is a local minimum with respect to $P_{0}$ (equilibrium). In
the presence of an external electric field $\delta\boldsymbol{E}$
the ferroelectric acquires an additional average (parallel) polarization
$\delta\boldsymbol{P}$. We assume small variations $\delta P$ and
$\delta E$, so that the variation of the (average) energy density
is 
\begin{equation}
\delta\mathcal{E}=a\delta P^{2}-2b\delta P\delta E+c\delta E^{2}>0\,\,\,,\label{1}
\end{equation}
where $a,b,c$ are some coefficients (first-order variation is zero
at equilibrium). In general, these coefficients depend on ferroelectric
and temperature. Small temperature variations $\delta T$ in the coefficients
$a$, $b$, $c$ make the above energy variation a higher-order expression,
so, in the first approximation, we neglect the temperature dependence
of these coefficients. The energy $\delta E^{2}/8\pi$ of the external
field can be included in equation (\ref{1}), so that the ferroelectric
plus the external field becomes an isolated system. The energy density
can be written as 
\begin{equation}
\delta\mathcal{E}=a\left(\delta P-\frac{b}{a}\delta E\right)^{2}+\left(c-\frac{b^{2}}{a}\right)\delta E^{2}\,\,\,,\label{2}
\end{equation}
where $a>0$ and $ac>b^{2}$. It is convenient to change the notations
according to $\delta P\rightarrow P$ and $\delta E\rightarrow E$,
and introduce the parameters $\lambda=b/a$ and $E_{c}$, according
to $\delta\mathcal{E}=\left(c-a\lambda^{2}\right)E_{c}^{2}$; the
above equation becomes 
\begin{equation}
a\left(P-\lambda E\right)^{2}+\left(c-a\lambda^{2}\right)E^{2}=\left(c-a\lambda^{2}\right)E_{c}^{2}\,\,\,,\label{3}
\end{equation}
 where $c>a\lambda^{2}$. This energy is conserved, $\left(c-a\lambda^{2}\right)E_{c}^{2}$
is constant, $P$ and $E$ are variables, and equation (\ref{3})
gives 
\begin{equation}
P_{\pm}=\lambda E\pm\sqrt{\frac{c-a\lambda^{2}}{a}}\sqrt{E_{c}^{2}-E^{2}}\,\,.\label{4}
\end{equation}
 The functions $P_{\pm}(E)$ can be viewed as two branches of a simplified
polarization hysteresis, as shown in Fig. \ref{fig:Fig.1}. The remanent
polarization ($P_{+}(E=0)$) is 
\begin{equation}
P_{r}=\sqrt{\frac{c-a\lambda^{2}}{a}}E_{c}\label{5}
\end{equation}
 and equation (\ref{4}) can be re-written as 
\begin{equation}
P_{\pm}=\lambda E\pm P_{r}\sqrt{1-E^{2}/E_{c}^{2}}\,\:;\label{6}
\end{equation}
the coercive field ($P_{-}(E_{coerc})=0$) is given by 
\begin{equation}
E_{coerc}=\sqrt{1-a\lambda^{2}/c}E_{c}=\sqrt{1-b^{2}/ac}E_{c}\,\,.\label{7}
\end{equation}

The field varies between $-E_{c}$ and $+E_{c}$, $P_{\pm}$ are symmetrical
with respect to inversion in the $E,P$-plane, and $P_{+}$ has a
maximum $P_{s}=\sqrt{c/a}E_{c}$ (for $E=\sqrt{a/c}\lambda E_{c}$),
which may be taken as the saturation value of the polarization; similarly,
$P_{-}$ has a minimum $-P_{s}$ for $E=-\sqrt{a/c}\lambda E_{c}$.
According to equation (\ref{6}), the inversion symmetry means 
\begin{equation}
P_{+}(E)=-P_{-}(-E)\,\,.\label{8}
\end{equation}
From the above equations the ratios $c/a$ and $\lambda=b/a$ are
given by 
\begin{equation}
\begin{array}{c}
c/a-\lambda^{2}=\left(E_{coerc}P_{s}/E_{c}^{2}\right)^{2}=\left(P_{r}/E_{c}\right)^{2}\,\,,\\
\\
\lambda^{2}=\left(P_{s}^{2}-P_{r}^{2}\right)/E_{c}^{2}\,\,.
\end{array}\label{9}
\end{equation}
The first equation (\ref{9}) is a consistency relation, which can
be used to check the validity of the model. The model is valid for
$E\ll E_{c}$ and $(c-a\lambda^{2})E_{c}^{2}$ much smaller than the
local energy density. For example, $E_{c}=30kV/mm$, $P_{s}=15\mu C/cm^{2}$,
$P_{r}=3\mu C/cm^{2}$ and $E_{coerc}=5kV/cm$ for $Ba_{0.87}Sr_{0.13}TiO_{3}$
(room temperature)\cite{key-11} do not satisfy the first equation
(\ref{9}), because of the large values of $E_{c}$ ($1statvolt/cm=3\times10^{4}V/m$,
$1C=3\times10^{9}statcoulomb$). For small values of $E$ and $E_{c}$
the model provides a satisfactory parametrization of the hysteresis
curves. For $E\ll E_{c}$ we may use $P_{\pm}\simeq\lambda E\pm P_{r}$,
$E_{coerc}\simeq P_{r}/\lambda$ and $P_{s}\simeq\lambda E_{c}+P_{r}$,
and the first equation (\ref{9}) is satisfied. In the presence of
an external electric field $E$ the work $-PdE<0$ should be minimal.
Therefore, the paths from $\pm E_{c}$ to $\pm E_{coerc}$ along the
$P_{\pm}$-branches are unstable ($-PdE>0$), so we expect a fall
from $P_{s}$ (or $\lambda E_{c}$) to zero ($E_{coerc})$, or a jump
from $-P_{s}$ (or $-\lambda E_{c}$) to zero ($-E_{coerc}$), in
qualitative accordance with the experimental sigmoid shape of the
hysteresis curves. 

Similarly, we have a variation of the entropy density, which is a
series expansion in powers of $\delta P$ and $\delta E$. Part of
this variation is $\delta S_{0}=\delta\mathcal{E}/T$, where $T$
is the temperature; it is a constant. As long as the temperature variation
of the coefficients is neglected, both the constant energy $\delta\mathcal{E}$
and entropy $\delta S_{0}$ may be neglected. The other part of the
entropy variation includes only powers of $\delta P$, because the
external field is not a thermodynamic system. The inversion symmetry
requires this entropy expansion to begin with $\delta P^{2}$. In
general, the number of available states is proportional to $(\boldsymbol{P}_{0}+\delta\boldsymbol{P})^{2}$,
and the average over the directions of $\boldsymbol{P}_{0}$ shows
that the entropy should increase, as expected for an isolated system.
Therefore, we may write this part of the entropy variation as 
\begin{equation}
\delta S=dP^{2}\,\,\,,\label{10}
\end{equation}
 where $d>0$ is a positive constant (whose temperature dependence
may be neglected). 

\section{Hysteresis cooling}

Let us consider a thermally isolated ferroelectric at constant pressure
in the presence of an external electric field $E$ (also, we may neglect
the volume changes). Let us assume that the field increases from $E=0$,
the ferroelectric being either at point $A$ or at point $B$ in Fig.
\ref{fig:Fig.1}. The energy variation of the ferroelectric in the
field is $-PdE$; it is minimal for $P>0$, $dE>0$ (or $P<0$, $dE<0$).
This is a decrease of the ferroelectric energy, on account of the
work done by the ferroelectric upon the charges which generate the
external field. Since $P$ has the same sign as $dE$, the ferroelectric
state corresponds to point $A$ and branch $P_{+}$ in Fig. \ref{fig:Fig.1}.
If the ferroelectric is initially at point $B$, the polarization
jumps to point $A$ and follows the branch $P_{+}$, in order to minimize
the energy. We note that the two polarization branches correspond
to metastable states, and the thermodynamic states are defined by
thermodynamic variables, like pressure, temperature, electric field
and, in addition, the polarization branch. The electric field varies
up to $E_{0}\ll E_{c},$(in fact, $E_{0}\ll E_{coerc})$ at point
$1$ in Fig. \ref{fig:Fig.1}, and the ferroelectric state follows
the branch $P_{+}$ up to that point, where the electric field begins
to decrease. At that point the ferroelectric jumps to point $2$ and
follows the branch $P_{-}$ up to point $3$, where the electric field
reaches its minimum value $-E_{0}$. Another jump occurs from point
$3$ to $4$, and the ferroelectric follows the branch $P_{+}$ up
to the origin, closing, clockwise, a cycle (loop). We note that the
states $1$ and $3$ are equivalent, and so are the states $2$ and
$4$, due to the inversion symmetry. Consequently, the cycle $4-1-2-3-4$
is in fact the path $4-1$ (or $2-3$) taken twice. Also, we note
that $P_{0}$ does not contribute to the cycle. By making use of equation
(\ref{6}), the work done along the hysteresis cycle is given by 
\begin{equation}
\begin{array}{c}
W=-\int_{-E_{0}}^{E_{0}}P_{+}dE-\int_{E_{0}}^{-E_{0}}P_{-}dE=\\
\\
=-\int_{-E_{0}}^{E_{0}}\left(\lambda E+P_{r}\sqrt{1-E^{2}/E_{c}^{2}}\right)dE-\\
\\
-\int_{E_{0}}^{-E_{0}}\left(\lambda E-P_{r}\sqrt{1-E^{2}/E_{c}^{2}}\right)dE=\\
\\
=-2\int_{-E_{0}}^{E_{0}}\left(\lambda E+P_{r}\sqrt{1-E^{2}/E_{c}^{2}}\right)dE\simeq-4E_{0}P_{r}\,\,.
\end{array}\label{11}
\end{equation}

We can see that the work has negative values. Since the ferroelectric
plus the field is an isolated system, the energy is conserved. On
the other hand, the energy associated to the polarization $P$, given
by equation (\ref{1}), is constant. It follows that the internal
energy of the ferroelectric varies by $\Delta\mathcal{E}=W<0$, on
account of the work done upon the external charges. Also, we have
a variation of the entropy $\Delta S$, so that we may write 
\begin{equation}
c_{p}\Delta T=T\Delta S=\Delta\mathcal{E}=W<0\,\,\,,\label{12}
\end{equation}
 where $c_{p}$ is the specific heat at constant pressure (and constant
field) and $T$ is the temperature. We can see that we have a temperature
decrease $\Delta T=W/c_{p}$. This is a cooling effect. The ferroelectric
is capable of absorbing heat, from a warm body, for example in an
isothermal process. 

Also, along the hysteresis cycle $4-1-2-3-4$ we have a variation
$\Delta(\delta S)$ of the entropy associated with the polarization,
given by equation (\ref{10}), 
\begin{equation}
\begin{array}{c}
\Delta(\delta S)=d\Delta(P^{2})=2d\left[P_{+}^{2}(E_{0})-P_{+}^{2}(-E_{0})\right]=\\
\\
=8d\lambda E_{0}P_{r}\sqrt{1-E_{0}^{2}/E_{c}^{2}}\simeq8d\lambda E_{0}P_{r}=-2d\lambda W>0\,\,.
\end{array}\label{13}
\end{equation}
 Since the ferroelectric is isolated, the total entropy should increase,
and the process is irreversible. Therefore, 
\begin{equation}
\Delta S+\Delta(\delta S)=\frac{W}{T}+\Delta(\delta S)=\left(-\frac{1}{2d\lambda T}+1\right)\Delta(\delta S)>0\,\,\,,\label{14}
\end{equation}
 which implies $2d\lambda T>1$ ($\Delta(\delta S)>0$). Indeed, we
note that the cycle cannot be taken in the reverse direction, because
the polarization $P$ and the variation $dE$ of the electric field
should be parallel, in order to minimize the energy. 

Also, we can consider the ferroelectric in thermal contact with a
warm body, exchanging with it a heat $T\Delta S_{th}$ at constant
temperature (an isothermal process), which is transformed in work,
$T\Delta S_{th}=-W>0$ ($\Delta S_{th}=-\Delta S$). This is an irreversible
process, the additional entropy variation $\Delta(\delta S)>0$ being
positive. 

If we take into account the temperature variation of the polarization,
the (total) energy variation is given by
\begin{equation}
\left(\frac{\partial\mathcal{E}}{\partial E}\right)_{T}=T\left(\frac{\partial S}{\partial E}\right)_{T}-P=T\left(\frac{\partial P}{\partial T}\right)_{E}-P\,\,\,,\label{15}
\end{equation}
 where $P$ is the total polarization (including $P_{0}$). We can
see that the energy variation includes the term $T\left(\partial P/\partial T\right)_{E}dE$,
besides $-PdE$ used above ($P_{0}$ does not contribute to the hysteresis
cycle). In equation (\ref{15}) the entropy variation can be estimated
from $dS/dE=P/T$ (at constant energy), which leads to $\Delta S=-W/T>0$
and $\Delta S=\Delta S_{th}$, given above (for one cycle). Similarly,
\begin{equation}
\begin{array}{c}
\left(\frac{\partial T}{\partial E}\right)_{\mathcal{E}}=\frac{\partial(T,\mathcal{E})}{\partial(E,\mathcal{E})}=\frac{\partial(T,\mathcal{E})/\partial(E,T)}{\partial(E,\mathcal{E})/\partial(E,T)}=-\frac{\left(\partial\mathcal{E}/\partial E\right)_{T}}{\left(\partial\mathcal{E}/\partial T\right)_{E}}=\\
\\
=\frac{-T\left(\partial\mathcal{S}/\partial E\right)_{T}+P}{\left(\partial\mathcal{E}/\partial T\right)_{E}}=\frac{-T\left(\partial P/\partial T\right)_{E}+P}{\left(\partial\mathcal{E}/\partial T\right)_{E}}=-\frac{1}{c_{p}}\left(T\left(\frac{\partial P}{\partial T}\right)_{E}-P\right)\,\,\,,
\end{array}\label{16}
\end{equation}
which is the same as equation (\ref{15}), since $TdS-PdE=c_{p}dT+d\mathcal{E}=0$.
The above formulae are similar with those describing the Joule-Thomson
process.\cite{key-12} 

The temperature dependence of the hysteresis loops is not yet conclusively
established (see, for example, Refs. \cite{key-13,key-14} and References
therein). In general, for sufficiently low temperature below the critical
temperature, where the domains are well developed, the temperature
dependence of the hysteresis curves is weak. According to our discussion
above, $P$ is a variation of $P_{0}$. As a model calculation we
may use $P_{0}\sim P_{0}^{*}\left(1-T/T_{c}\right)_{0}^{1/2}$, where
$P_{0}^{*}$ is the maximum polarization and $T_{c}$ is the critical
temperature. It follows that $P=P^{*}\left(1-T/T_{c}\right)^{1/2}$,
where $P^{*}$ is given by equation (\ref{6}). For $T/T_{c}\ll1$
the term $T\left(\partial P/\partial T\right)_{E}$ in equation (\ref{15})
is approximately $-(T/2T_{c})\left(P_{0}^{*}+P^{*}\right)$, which
may be neglected in that equation ($P_{0}^{*}$ does not contribute
to the cycle). 

For a numerical illustration we use data for $BaTiO_{3}$-based materials,
with critical temperature in the region of the room temperature.\cite{key-15}-\cite{key-18}
For a modest value $P_{r}=10^{2}statvolt/cm$ ($\simeq0.03\mu C/cm^{2}$,
$3\times10^{6}V/m$) and $E_{0}=0.1statvolt/cm$ (3$\times10^{3}V/m$),
we get $W=-40erg/cm^{3}$. For a density $6g/cm^{3}$ and a specific
heat $5\times10^{6}erg/g\cdot K$ the temperature variation is $\Delta T=-1.3\times10^{-6}K$.
If the external electric field is oscillating with frequency $\omega$,
we have an energy rate $-\frac{2\omega}{\pi}E_{0}P_{r}$, and $\omega=10MHz$
leads to $\simeq-2K/s$, which is a very efficient cooling (though
high frequencies may distort the hysteresis curves\cite{key-19}).
We note that the energy density stored by the ferroelectric is of
the order of the polarization squared, and we may take the polarization
of the order $10^{4}statvolt/cm$. The extracted energy should be
continuously replaced by the energy taken up from the warm body, and
the actual efficiency of the process depend on the thermoconduction
coupling of the ferroelectric and the warm body. It may happen that
after a prolonged cycling the ferroelectric gets \textquotedbl frozen\textquotedbl ,
and it needs to be heated up in order to be reusable. The results
are similar for a ferromagnet, with $E$ and $P$ replaced by the
external magnetic field $H$ and magnetization $M$. Data for ferromagnets
can be found in Refs. \cite{key-20}-\cite{key-22}.

\section{Discussion}

We note that the process described above is distinct from the electrocaloric
process along the hysteresis cycle, where an increase of temperature
is produced. The cooling process transforms heat in work, while in
the heating process that implies the electrocaloric effect work is
transformed in heat. 

It is interesting to compare the above results with the electrocaloric
effect along the hysteresis loop. As it is well known, the temperature
variation in the electrocaloric effect is\cite{key-23} 
\begin{equation}
dT=-\frac{T}{c_{p}}\left(\partial P/\partial T\right)_{E}dE\,\,.\label{17}
\end{equation}

By using $P=P^{*}\left(1-T/T_{c}\right)^{1/2}$, with $P^{*}$ given
by equation (\ref{6}), the temperature variation can be estimated
as 
\begin{equation}
\begin{array}{c}
\Delta T\simeq\frac{T}{c_{p}T_{c}}\int_{-E_{0}}^{E_{0}}P^{*}dE=\\
\\
=\frac{T}{c_{p}T_{c}}\int_{-E_{0}}^{E_{0}}\left(\lambda E+P_{r}\sqrt{1-E^{2}/E_{c}^{2}}\right)dE\simeq\frac{2E_{0}P_{r}T}{c_{p}T_{c}}\,\,.
\end{array}\label{18}
\end{equation}
For $P_{r}=10^{2}statvolt/cm$, $E_{0}=0.1statvolt/cm$, and $c_{p}=3\times10^{7}erg/cm^{3}\cdot K$,
we get $\Delta T\simeq0.7(T/T_{c})\times10^{-6}K$. We note the sign
of $\Delta T$ ($>0$). 

The distinction between the cooling effect and the heating (electrocaloric)
effect is governed by the rate of variation of the external field.
If the field is changed sufficiently fast, the work is $-PdE$, where
the polarization $P$ follows the field $E$ along the hysteresis
curve. This is the cooling effect. If the field is changed slowly,
it may react back upon the ferroelectric, which receives an amount
of energy equal to the variation of $PdE$. This is the heating effect.
The energy variation $\Delta PdE$ is equal to the heat $-\Delta TdS$,
\emph{i.e.} $\Delta PdE=-\Delta T\left(c_{p}dT\right)/T$, which is
equation (\ref{17}) (the variations $\Delta P$ and $dE$ are independent).
This is a slower process, which needs a finite time $\tau$ to thermalize
the excess of energy. Moreover, the preceding equation can be written
as $(\Delta P/\Delta T)_{E}dE+\left(\partial S/\partial T\right)_{E}dT=0$,
or $(\partial S/\partial E)_{T}dE+\left(\partial S/\partial T\right)_{E}dT=0$,
which is the adiabatic condition $dS=0$. Therefore, this electrocaloric
process looks formally as an adiabatic, reversible process, as it
is well known. However, we cannot travel through the hysteresis cycle
in the reverse direction, since the polarization must be parallel
to the field variation. Actually, we have the additional entropy variation
$\delta S>0$ (equation (\ref{10})), specific to hysteresis, which
tells that the process is, in fact, irreversible.

Time $\tau$ is a thermalization time of the domains. We can have
an estimate of this (average) time by using $\tau\simeq\sqrt{ma^{2}/T}$,
where $m$ is the domain mass and $a$ is of the order of the domain
dimension. For $T=300K$, $a=1\mu m$ ($10^{-4}cm$) and $m=4\times10^{-12}g$
(density $4g/cm^{3}$) we get $\tau\simeq1ms$. (We note that this
time is much shorter than the relaxation time associated with an activation
energy, see, for example, Refs. \cite{key-24,key-25}). We may say
that if we reverse the external field in less than $1ms$, we get
the cooling effect (\emph{e.g.}, with the above $10MHz$ frequency);
if we change the field slower, in a time longer than $1ms$, we get
the heating (electrocaloric) effect. Indeed, hysteresis heating is
observed for low switching frequencies.\cite{key-26,key-27} 

In conclusion, we identified a cooling process by means of an external
field along hysteresis loops, and used a simplified hysteresis model
in order to illustrate it. The model is based on the energy variation
of a ferroelectric or a ferromagnet in an external field. For comparison
we discussed also the electrocaloric effect along hysteresis loops. 

\textbf{Acknowledgements.} The author is indebted to the members of
the Laboratory of Theoretical Physics at Magurele, especially, to
dr L. C. Cune, for many valuable discussions. The work was carried
out within the Program Nucleu, funded by the Romanian Ministry of
Research, Innovation and Digitization, project no. PN23210101/2025.
The author declares no competing interest. All data have been taken
from the cited literature.

\textbf{Conflict of interest.} The author has no conflicts to disclose.

\textbf{Data availability.} Data sharing is not applicable to this
article as no new data were created or analyzed in this study.

\textbf{Author Contributions.} M. Apostol: Conceptualization, analysis,
investigation, methodology, writing.

\end{document}